# A Nanostructual Microwave Probe Used for Atomic Force Microscope

Y. Ju*, M. Hamada*, T. Kobayashi**, H. Soyama**
\* Department of Mechanical Science and Engineering, Nagoya University
Furo-cho, Chikusa-ku, Nagoya 464-8603, Japan
\*\* Department of Nanomechanics, Tohoku University
Aoba 6-6-01, Aramaki, Aoba-ku, Sendai 980-8579, Japan

*Abstract-* In order to develop a new structure microwave probe, the fabrication of AFM probe on the GaAs wafer was studied. A waveguide was introduced by evaporating Au film on the top and bottom surfaces of the GaAs AFM probe. A tip having 8 μm high, and curvature radius about 50 nm was formed. The dimensions of the cantilever are 250×30×15 μm. The open structure of the waveguide at the tip of the probe was introduced by using FIB fabrication. AFM topography of a grating sample was measured by using the fabricated GaAs microwave probe. The fabricated probe was found having nanometer scale resolution, and microwave emission was detected successfully at the tip of the probe by approaching Cr-V steel and Au wire samples.

## I. INTRODUCTION

With the development of nanotechnology, the measurement of electrical properties in local area of materials and devices has become a great need. Although a lot kinds of scanning probe microscopes have been developed for satisfying the requirement of nanotechnology, a microscope technique which can determine electrical properties in local area is still under developing. Recently, microwave microscope has been an interest to many researchers [1-4], due to its potential in the evaluation of electrical properties of materials and devices. The advance of microwave is that the response of materials is directly relative to the electromagnetic properties of materials. Recently, we have proposed a microwave atomic force microscope (M-AFM) which is expected to be able to realize the evaluation of electrical properties as well as the measurement of topography of materials in nanoscale orders [5]. This technique combined the characteristic of the microwave microscope and the atomic force microscope which has nanoscale spatial resolution and can keep standoff distance constantly by the atomic force between the tip and sample. The details to fabricate the M-AFM probe using wet etching technique have been described in Ju et al. [5], and the evaluation of the capability to sense the topography of materials by using the M-AFM probe was demonstrated in Ju et al. [6]. In this paper, the propagation of microwave signal in the probe and the emission of the signal at the tip of the probe were carried out.

## II. PROBE FABRICATION

In this research, no doped GaAs wafer was used as the substrate of the probe in order to restrain the attenuation of microwave propagating in the probe. To obtain the desired structure, wet etching was used to fabricate the probe. In contrast to dry etching, a side etching occurs under the etching mask. Utilizing this property, a cone-shaped microtip can be obtained. Early studies suggested that a square resist pattern having 14 μm sides and one side at 45° to the <011> direction is the most suitable mask for etching the tip of the probe [7]. By considering the chemical activities at different crystalline planes, the length direction of the etching mask for forming the beam of the cantilever was patterned along the <011> direction. Consequently, side-etching occurs under the resist mask, and mesa-type planes appear at both sides of the beam (45° inclined plane). On the other hand, an inverse-mesa type plane is formed at the end of the beam (60°–75° inclined plane) [6]. Under the same conditions as the beam fabrication process, the body of the probe was formed by backside etching. The etching mask was patterned on the bottom surface, and etching was carried out until the substrate was penetrated. After that, Au film of 50 nm thickness was deposited on the top and bottom surfaces of the probe to

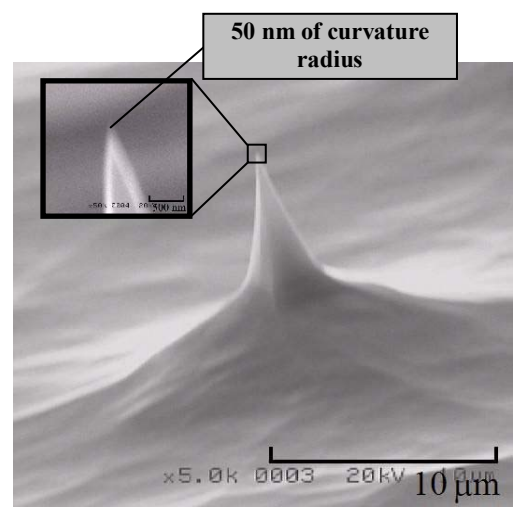

Fig. 1. SEM photograph of the tip of the GaAs AFM probe.

III. EXPERIMENTAL RESULTS

A. *AFM topography*

In order to confirm the spatial resolution of the fabricated M-AFM probe, the AFM topography of a grating sample having 2000 line/mm was measured by using the fabricated probe. JSPM-5400 was used for measuring the sample in non-contact mode which is suitable for carrying out microwave measurement. The scan area was 2 × 2 μm, scan speed was 3.0 μm/sec. Fig. 4 shows the AFM topography of the grating sample obtained by the M-AFM probe. Fig. 4a) is 2-D image of the topography, and Fig. 4b) shows the 3-D image. The white spots in these figures are due to micro-dusts on the sample surface. It is observed that the fabricated M-AFM probe has the capability to catch the AFM topography and height information in nanometer orders.

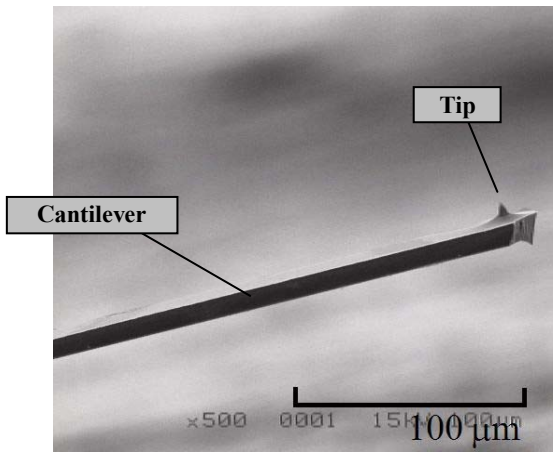

Fig. 2. The cantilever of the probe having dimensions of 250×30×15 μm.

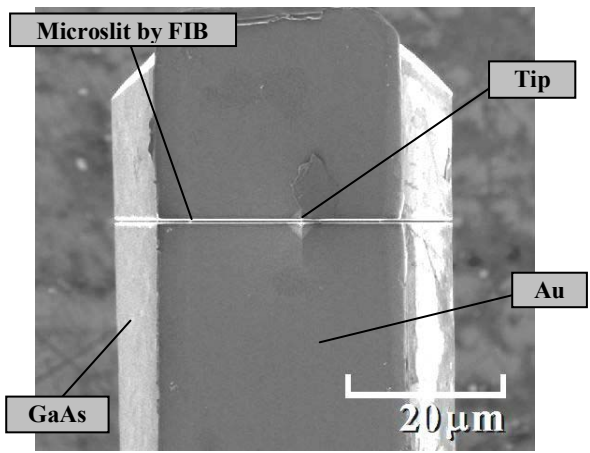

Fig. 3. SEM photograph of the microslit of the microwave AFM probe introduced by FIB fabrication.

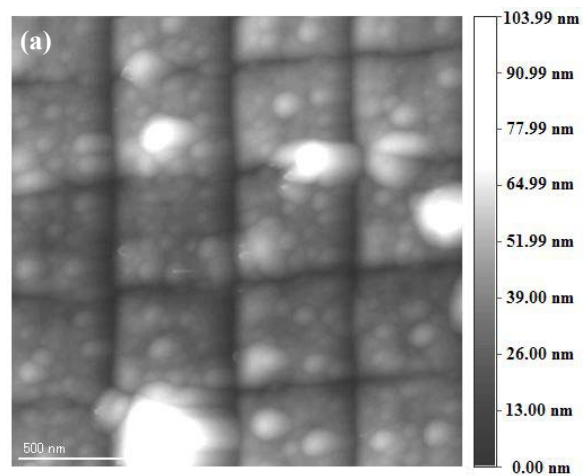

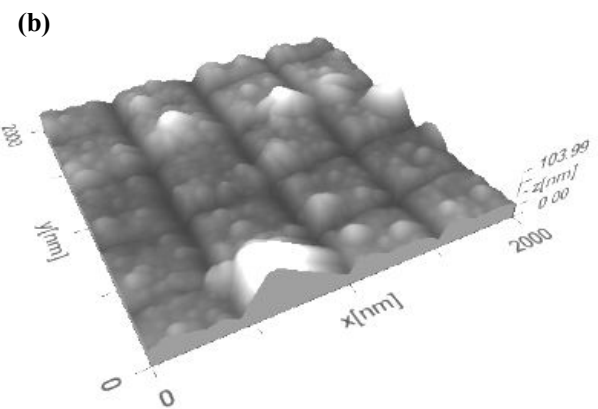

Fig. 4. AFM surface topography of the grating sample obtained by the fabricated M-AFM probe:
(a) 2-D image, (b) 3-D image.

propagate a microwave signal in the probe. Both plane surfaces of the waveguide, which were made of evaporated Au film, were connected at the end of the beam. However, there was no Au film on the sides of the beam, since the formed inclined planes at the beam sides were not facing the direction of evaporation. By using focused ion beam (FIB) fabrication, a slit at the tip of the probe was formed to open the connection of the Au film on the two surfaces of the probe. Consequently, the microwave will propagate along the probe and emit at the tip of the probe.

Scanning electron microscopy (SEM) observation was carried out for the fabricated M-AFM probe. As shown Fig. 1, a sharp tip having high aspect ratio (2.0) was obtained. The tip is the 8 μm high, and the curvature radius of the tip is about 50 nm. Fig. 2 shows the observation of the cantilever of the probe, where the tip is located near the front edge of the cantilever. As shown in Fig. 3, a microslit was introduced cross the cantilever through the center of the tip by FIB fabrication. The width of the microslit is small than 100 nm.

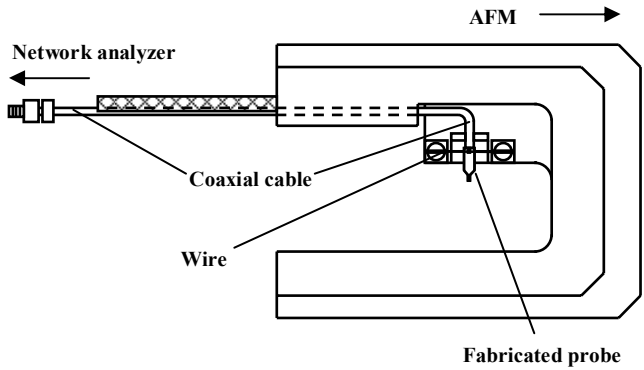

Fig. 5. The schematic diagram of the improved probe holder.

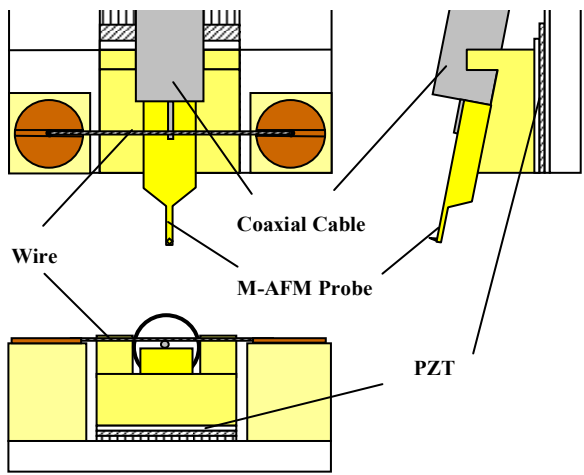

Fig. 6. The connection of the probe with the coaxial line.

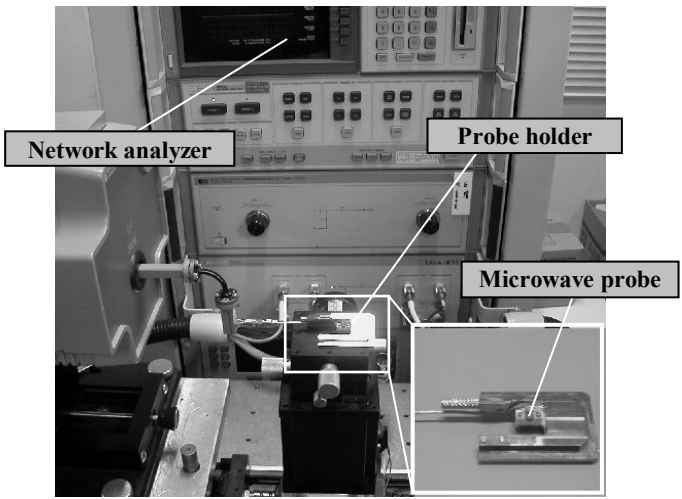

Fig. 7. Photograph of microwave measurement system.

### B. Propagation and detection of microwave

The microwave signal was measured by a network analyzer where the working frequency is 75~110 GHz. A coaxial line was used to connect the M-AFM probe with the network analyzer. In order to realize such connection, the coaxial line having the diameter of 1 mm was fixed on the probe holder which is used to set up an AFM probe for AFM measurement. The schematic diagram of the improved probe holder is shown in Fig. 5, and the enlarged view of the probe connection is shown in Fig. 6. The outer and inner conductors of the coaxial line are connected to the bottom and top surfaces of the M-AFM probe, respectively. Therefore, microwave transmission line changes form the coaxial line to the parallel plate waveguide (in the probe). By using this holder, M-AFM probe can be set up repeatedly, and the measurement of microwave response and AFM surface profile can be realized simultaneously.

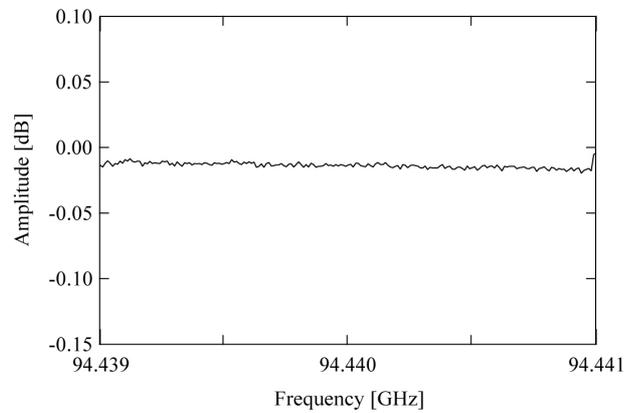

Fig. 8. The measurement result of the amplitude of reflection coefficient when the test sample is absent at the tip of the M-AFM probe.

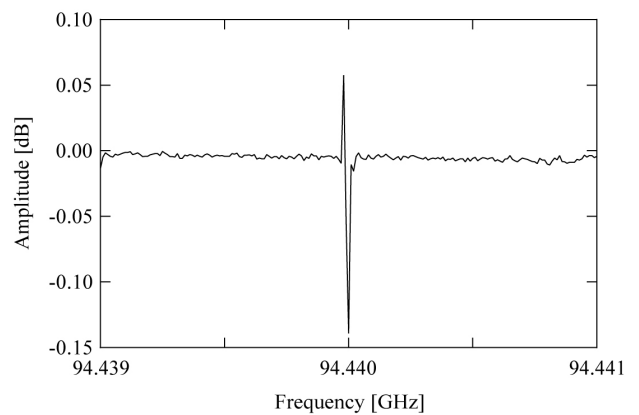

Fig. 9. The measurement result of the amplitude of reflection coefficient when the Cr-V steel sample is present at the tip of the M-AFM probe.

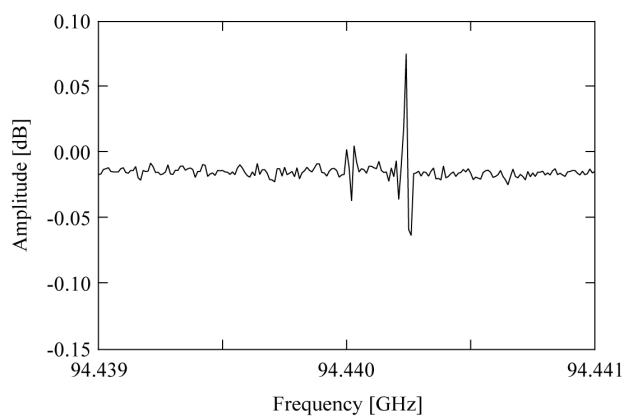

Fig. 10. The measurement result of the amplitude of reflection coefficient when the Au wire is present at the tip of the M-AFM probe.

Microwave propagation should perform accurately in the fabricated M-AFM probe in order to evaluate electrical property of materials. To confirm that, the response of microwave reflection was examined by approaching a Cr-V steel sample and an Au wire (50 μm diameter) at the tip of the probe. The microwave measurement system used in the experiment is shown in Fig. 7. The experiment was carried out as follows: microwave signal generated in network analyzer was feed to the M-AFM probe; the test samples were approached to the tip of the probe; the amplitude-frequency response was measured. The reflection signal will change due to the difference between characteristic impedance of the samples and air when microwave is propagating in the probe and emitting at the tip. It means that microwave emission at the tip of the M-AFM can be evaluated by measuring the response of the microwave signal.

Fig. 8 and Fig. 9 show the measurement results of the microwave response. The amplitude of the reflection coefficient was measured by sweeping the working frequency. Here, the swept frequency range is 2 MHz and the center frequency is 94.44 GHz. Fig. 8 is the result that the test sample is absent at the tip of the M-AFM probe. Fig. 9 is the result that the Cr-V steel sample is present at the tip of the probe. When the test sample is made to be approached at the tip of the probe, as shown in Fig. 9, the peak changing in amplitude is observed clearly. This result indicates that the microwave emitted at the tip of the probe. The amount of the variation of the amplitude is about 0.2 dB. Fig. 10 is the result that the Au wire is present at the tip of the probe. It is confirmed that the amount of variation of the amplitude is about 0.15 dB. Here, since the Au wire has the diameter of 50 μm which is great small than the dimensions of Cr-V steel sample, a smaller signal was obtained. However, it should be noted that the measured signal is sensitive enough for the evaluation of electrical property in the microwave AFM measurement.

## IV. CONCLUSION

M-AFM probes were fabricated on the GaAs wafer by using wet etching process. A waveguide was introducing on the probe by evaporating Au film on the both surfaces of the probe. SEM observation shows that a tip having 8 μm high and curvature radius about 50 nm were formed. The open structure of the waveguide at the tip of the probe was obtained by using FIB fabrication. AFM measurement was performed. It is indicated that GaAs microwave probe has a capability to catch AFM topography of materials in nanometer scale. Microwave emission was detected at the tip of the probe which is large enough for the evaluation of electrical properties in the microwave AFM measurement.

ACKNOWLEDGMENT

This work was supported by the Japan Society for the Promotion of Science under Grant-in-Aid for Scientific Research (S) 18106003 and (A) 17206011; Ministry of Education, Culture, Sports, Science and Technology of Japan under Grant-in-Aid Exploratory Research 18656034.